\documentclass[1,12pt]{elsarticle}
\usepackage{graphicx}
\usepackage{graphicx,epstopdf}
\usepackage{caption}
\usepackage{subcaption}
\usepackage{float}
\usepackage{amssymb,amsmath}
\usepackage{exscale}
\usepackage{makeidx,shortvrb,latexsym}
\usepackage{epstopdf}
\usepackage{tabularx, booktabs, multirow}
\usepackage{array}
\usepackage[]{hyperref}
\usepackage{color}

\journal{International Journal of Fatigue, vol. 107, pp. 40-48, 2018.}

\newcounter{defcounter}
\begin{document}
\begin{frontmatter}

\title{Microstructure-based fatigue life model of metallic alloys with bilinear Coffin-Manson behavior}

\author{A. Cruzado$^{1, 2}$\corref{cor1}}
\author{S. Lucarini$^1$\corref{cor2}}
\author{J. LLorca$^{1, 3}$\corref{cor3}}
\author{J. Segurado$^{1, 3}$\corref{cor4}}
\address{
$^1$ IMDEA Materials Institute \\ C/ Eric Kandel 2, 28906, Getafe, Madrid, Spain. \\\ \\
$^2$ Department of Aerospace Engineering, Texas A$\&$M University\\  H. R. Bright Building, 701 Ross St, College Station, TX 77840, USA.\\\ \\
$^3$ Department of Materials Science, Polytechnic University of Madrid/Universidad Polit\'ecnica de Madrid \\ E. T. S. de Ingenieros de Caminos. 28040 - Madrid, Spain. 
}
\cortext[cor1]{Corresponding author at: IMDEA Materials Institute, Spain \\ E-mail address: aitor.cruzado@imdea.org} 

\begin{abstract}
A microstructure-based model is presented to predict  the fatigue life of polycrystalline metallic alloys which present a bilinear Coffin-Manson relationship. The model is based in the determination of the maximum value of a fatigue indicator parameter obtained from the plastic energy dissipated by cycle in the microstructure. The fatigue indicator parameter was obtained by means of the computational homogenization of a representative volume element of the microstructure using a crystal-plasticity finite element model. The microstructure-based model was applied to predict the low cyclic fatigue behavior of IN718 alloy at 400$^\circ$C  which exhibits a bilinear Coffin-Manson relationship under the assumption that this behavior is triggered by a transition from highly localized plasticity at low cyclic strain ranges to more homogeneous deformation at high cyclic strain ranges. The model predictions were in very good agreement with the experimental results for a wide range of cyclic strain ranges and two strain ratios ($R_\varepsilon$ = 0 and -1) and corroborated the initial hypothesis. Moreover, they  provided a micromechanical explanation for the influence of the strain ratio on the fatigue life at low cyclic strain ranges.
\end{abstract}


\end{frontmatter}

\section{Introduction}

The Coffin-Manson (C-M)  \citep{S91} phenomenological model is widely used to characterize the fatigue life of metallic alloys under strain-controlled cyclic deformation. In this model, the link between the fatigue life $N_f$ and the cyclic plastic strain amplitude, $\Delta\varepsilon_p$/2, is expressed as
\begin{equation}
\label{eq_(1):CF}
  \Delta \varepsilon_p/2 = C N_f^c
\end{equation}
\noindent where $C$ and $c$ are material constants that can be determined from experiments. This expression leads to a linear relationship  between $\Delta \varepsilon_p$/2 and $N_f$ in a log-log plot  and the slope is the exponent $c$, which is in the range -0.5 to -0.7 for many metals \citep{S91}. Further application of this model has shown that the connection between the fatigue life and the plastic strain amplitude is better expressed by a bilinear relationship \citep{R92}  in a number of metallic alloys including dual-phase steels \citep{MEDIRATTA1986}, Al alloys \cite{S88, PRASAD1994} and Ni-based superalloys \cite{LG85}, particularly IN718 \citep{BSP16, Xiao2005b, Sanders1981, Praveen2008}.

The origin of the bilinear relationship between $\Delta \varepsilon_p$/2 and $N_f$ has often been attributed to a change in the fracture mode triggered by the different magnitudes of the plastic strain amplitude. For instance, a transition from intergranular to transgranular fracture was reported by Prasad {\it et al.} \citep{PRASAD1994} in Al-Li alloys. In the particular case of IN718, the cyclic plastic strain at which the transition between the two Coffin-Manson regimes occurs is known to depend on the test temperature \citep{Sanders1981} and on the heat treatment \citep{Praveen2008} but the origin of this transition is still under debate. Sanders {\it et al.} \citep{Sanders1981} were the first  ones to analyze this behavior in a  range of temperatures and found  that the bilinear C-M behavior appeared below 423$^{\circ}$C. Transmission electron microscopy observations in these samples showed that the dominant deformation mechanisms were similar at low and high cyclic plastic strain amplitudes. They attributed the change of slope in the C-M law to a change from highly heterogeneous localized plastic slip at low cyclic strain amplitudes to more homogeneous deformation at higher cyclic strain amplitudes. Bhattacharya et al. \citep{BHATTACHARYYA1997} also reported a bilinear C-M law in IN718 tested at ambient temperature and  associated the dual slope to a  change in the dominant deformation from microtwinning at low cyclic strain amplitudes to slip band formation at high cyclic strain amplitudes. More recent fatigue tests at ambient temperature, together with  transmission electron microscopy observations, carried out by Praveen {\it et al.} \citep{Praveen2004, Praveen2008} in IN718 in the solution, peak-aged and over-aged conditions showed that the dominant deformation mechanism was dislocation slip in the whole cyclic strain amplitude range. The bilinear C-M was attributed, as in \citep{Sanders1981}, to the transition from highly localized plasticity at low cyclic strain amplitudes to more homogeneous deformation at high cyclic strain amplitudes. It should be noticed that crack propagation was predominately transgranular for both plastic strain amplitude ranges. 

The experimental determination of the parameters that define the bilinear C-M relationship is very expensive because it is necessary to determine the critical cyclic plastic strain amplitude at which the transition takes place as well as the slope of of C-M law in each region. Moreover, the particular election of the plastic strain as the fatigue indicator parameter does not seem to be the origin of this dual behavior. Some authors have used macroscopic indicators different to plastic strain, such as plastic dissipation per cycle \citep{Praveen2008}, but the results obtained are equivalent and two different regions were found when the number of cycles for crack nucleation is represented as a function of the parameter. An alternative approach is based in the application of microstructure-based fatigue life models based on the analysis of the cyclic deformation of the polycrystal by means of computational homogenization of a representative volume element of the microstructure \cite{Shenoy2007, Shenoy2008, McDowell2010, Cruzado2017}. These models can provide details of the evolution of the stress and strain fields for each slip system in each grain during cyclic deformation and this information can be used to predict the fatigue life following two different approaches. In the first one, introduced by Manonukul and Dunne \citep{Manonukul2004}, fatigue life is assumed to be controlled  by crack initiation \citep{Sweeney2012, Sweeney2013, Sweeney2015, Wan2016} which is related to the evolution of the cyclic stress and strain fields within the microstructure through the Fatigue Indicator Parameters (FIP). In the second approach, pioneered by Shenoy and McDowell \citep{Shenoy2007}, fatigue life is the sum of the number of cycles necessary for crack nucleation, microstructurally small crack growth  and crack propagation \citep{Shenoy2007,Musinski2012, Przybyla2013, Castelluccio2014}. While this latter approach is more rigorous from the physical viewpoint, it is difficult to correlate with experiments because of the difficulties associated with the detection of crack initiation or of the propagation of cracks whose length is of the order of the grain size.  

The application of these  strategies to fatigue is specially appealing for metallic alloys that present a bilinear C-M law because of their ability to naturally account for the heterogeneity of plastic deformation within the polycrystal, which stands at the origin of this behavior. This is the main objective of this investigation in which the low-cycle life of  IN718 alloy with a bilinear C-M behavior was modelled by means of microstructure-based fatigue life model. The constitutive equation for IN718 alloy under cyclic deformation as well as the cyclic jump strategy (to speed up calculations) were presented in a previous investigation \cite{Cruzado2017}. The focus of this paper is to relate the fatigue life with the microscopic FIP through a simple expression so the whole bilinear C-M behavior can be predicted. Thus, a simple -- but efficient -- combined experimental and simulation strategy can be used to predict the fatigue performance of materials with a bilinear C-M behavior. 

The paper is organized as follows. The experimental results of low cycle fatigue tests at $400^{\circ}$C in a fine grain IN718  alloy are presented in section 2. Section 3  summarizes the computational homogenization strategy to predict the cyclic behavior while the model for fatigue life prediction are developed and validated in section 4. The main conclusions of the work are found in section 5.
 
\section{Experimental behavior of IN718 alloy under low cycle fatigue}

The material studied in this work is wrought polycrystalline Inconel 718 superalloy, denominated  IN718 alloy, in final condition of solution and precipitation aged. It shows a fine microstructure with equiaxed $\gamma$ phase grains as the matrix, and a small fraction (about 12$\%$)  of $\delta$ phase (Ni3Nb) distributed preferentially at grain boundaries, and a minor amount (about 0.5$\%$) of second phase particles (niobium carbides, and titanium carbonitrides). As result of heat treatments material is strengthened by very fine $\gamma$' and $\gamma$'' precipitates within the grains, with diameters below 50 nm.
 
The specimens of IN718 present the same grain size in the surface and in the specimen interior. Moreover, the grain sizes observed in the longitudinal and transversal specimen surfaces were very similar. For this reason grains are assumed to be equiaxial and therefor the grain radii distribution completely characterizes the microstructure. This grain size was obtained evaluating transversal and longitudinal sections taken from specimens ends following ASTM E112 standard. The sample measured contained more than 1000 grains and the statistical treatment of data for the average and 99P/95CL bounds gave following results:   mean area 322 $\pm$ 22 $\mu$m$^2$, average equivalent circle 18.1 $\pm$ 0.6 $\mu$m, ASTM index, 8.6 $\pm$ 0.2.

With regard to mechanical test, uniaxial strain control low cycle fatigue tests were performed by ITP according to the standard ASTM E606-04  at 400$^{\circ}$C, on smooth cylindrical test pieces(5.08 mm diameter and of 12.7 mm gauge length) of fine grain size material.  Fatigue tests were carried out in a 100 kN load capacity MTS servo-hydraulic fatigue load frame under strain control. A trapezoidal wave form  was applied following the pattern 1 s (dwell) - 5 10$^{-3}$ s$^{-1}$ (ramp up) - 1 s (dwell) - 5 10$^{-3}$ s$^{-1}$ (ramp down).

The fatigue tests were carried out with cyclic strain ranges $\Delta \varepsilon /\Delta \varepsilon_{min}$ = 1, 1.25, 1.5, 1.75, 2.25, 2.75  with a strain ratio $R_\varepsilon$ = 0 and with strain ranges of $\Delta \varepsilon/\Delta \varepsilon_{min}$ = 1, 1.5,2, 2.5, 3 and 3.5 with $R_\varepsilon$ = -1. A normalization factor, $\Delta\varepsilon_{min}$, has been introduced for confidentiality reasons. The value of  $\Delta\varepsilon_{min}$,  corresponds to the smallest cyclic strain range used in the experimental campaign.

\begin{figure}[H]
\includegraphics[scale=0.90]{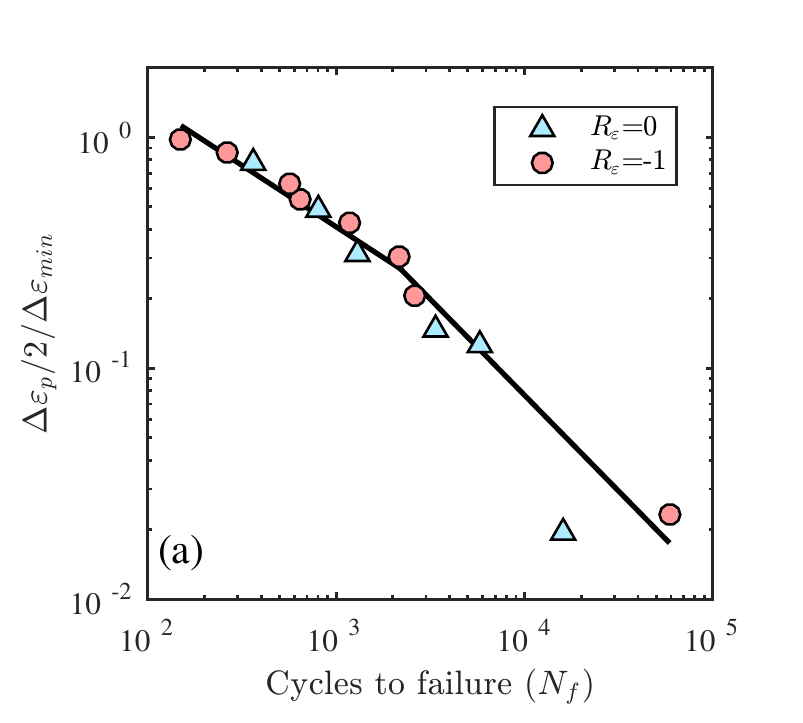}
\includegraphics[scale=0.90]{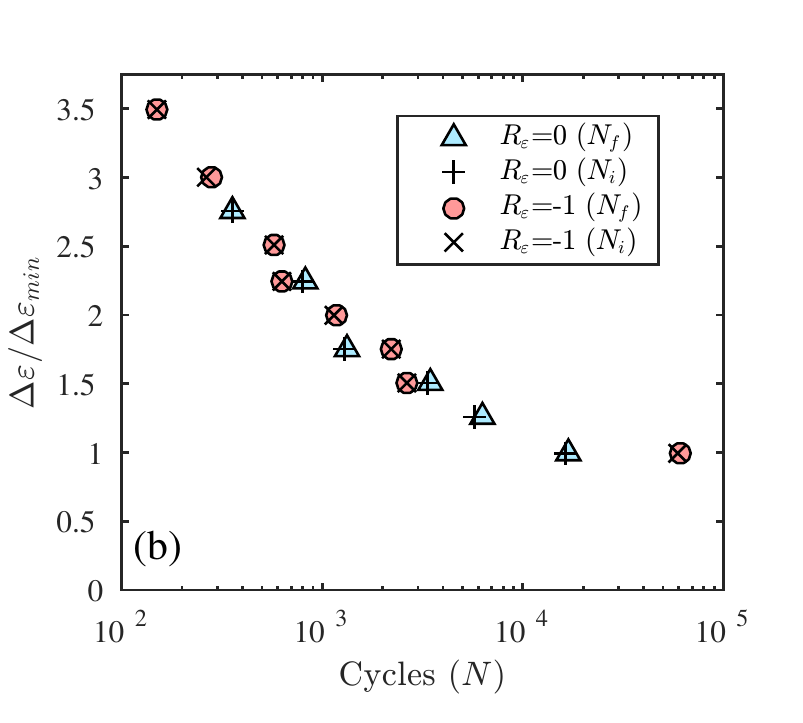}

\caption{(a) Cyclic plastic strain amplitude,$\Delta \varepsilon_p$/2, {\it vs.}  the number of cycles to failure, $N_f$, in IN718 alloy under fully-reversed ($R_\varepsilon$ = -1) and  non-symmetrical strain ratio ($R_\varepsilon$ = 0) at 400$^{\circ}$C. (b) Applied cyclic strain range, $\Delta\varepsilon$, {\it vs.} the number of cycles to failure $N_f$ . The number of cycles for crack initiation $N_i$ (5$\%$ load drop) are also included for comparison. Cyclic plastic strain amplitudes and strain ranges are normalized by $\Delta\varepsilon_{min}$. See text for details.} 

\label{Fig_1:LCF_test_data}
\end{figure}

The experimental results of the low cycle fatigue tests  are shown in Fig. \ref{Fig_1:LCF_test_data}(a), where the cyclic plastic strain amplitude is plotted {\it vs.} the number of fatigue cycles until failure. The alloy follows a bilinear C-M law and the transition between both slopes takes place for $N \approx$ 2000. It is worth noting that the slope for higher cyclic plastic strain amplitudes (lower fatigue life) is less pronounced. The same test results are represented in Fig. \ref{Fig_1:LCF_test_data}(b) in which the total applied cyclic strain range ($\Delta \varepsilon/\Delta \varepsilon_{min}$) is plotted as a function of the number of cycles to failure. This representation shows that the effect of strain ratio is small except for the smallest values of the cyclic strain range where the fatigue life with $R_\varepsilon = 0$ is four times shorter than with $R_\varepsilon = -1$.   An estimation of the number of cycles to crack initiation is also included in Fig.\ref{Fig_1:LCF_test_data}(b) together with the number of cycles to failure. Crack initiation was assumed when the maximum load during cyclic loading dropped by 5\%, as suggested in many studies \Citep{Brommesson2015, Wan2016}. From this estimation of the number of cycles for crack initiation, it can be concluded that the fatigue life was controlled by crack nucleation in the whole range of cyclic strain ranges covered in the experimental campaign. This result is in agreement with previous studies in IN718 alloy at 25$^{\circ}$C and 550$^{\circ}$C \citep{Fournier1977} in which the number of cycles to  form microcracks of  50 to 100 $\mu$m in length was measured through a marking technique based on oxidation.  It was found that  70\% of the life in low cycle fatigue tests was spent in crack initiation.  

\begin{figure}[H]
\includegraphics[scale=0.90]{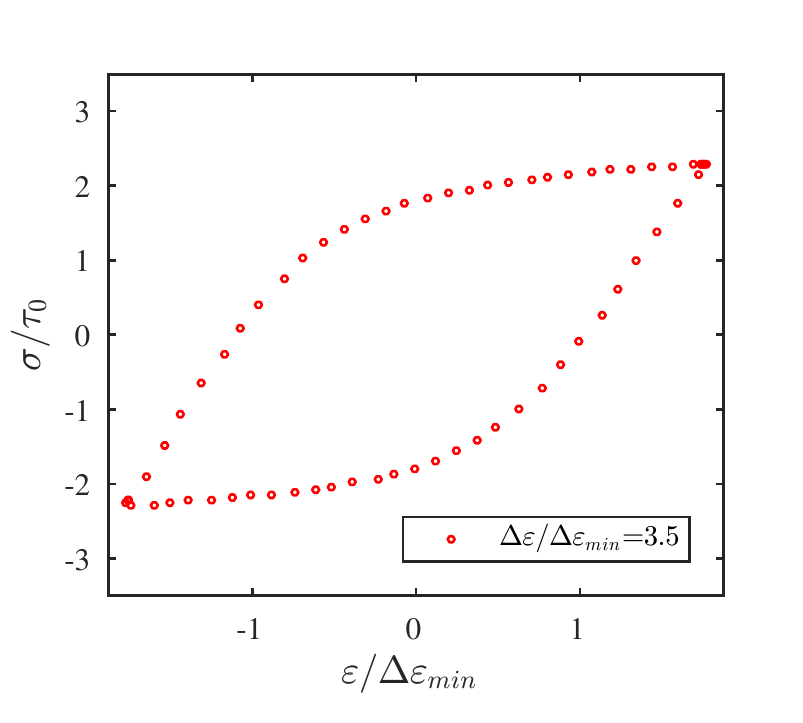}
\includegraphics[scale=0.90]{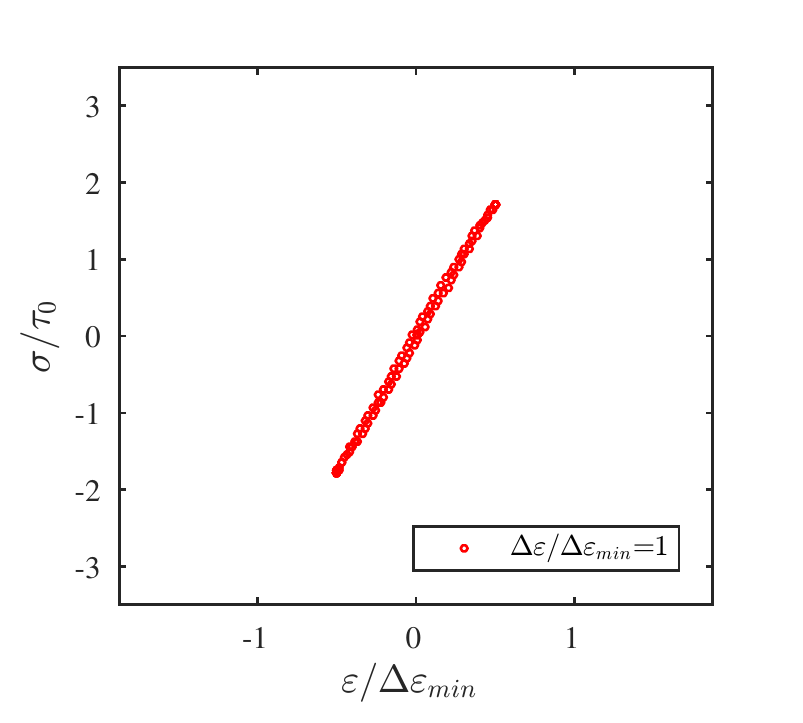}

\caption{Experimental results of the stabilized cyclic stress strain loops of IN718 at 400$^{\circ}$C tested under uniaxial tension with $R_\varepsilon$ = -1 for the two extreme strain ranges: (a)  for $\Delta \varepsilon/\Delta \varepsilon_{min}$=3.5 and (b) $\Delta \varepsilon/\Delta \varepsilon_{min}$=1.}
\label{stabilized cycles}
\end{figure}

The stress-strain hysteresis loop in the two extreme conditions ($\Delta \varepsilon/\Delta \varepsilon_{min}$=3.5 and 1),  are shown in Fig.\ref{stabilized cycles} for a strain ratio of $R_\varepsilon$ = -1.  The macroscopic cyclic behavior is mainly dominated by plastic deformation when $\Delta \varepsilon/\Delta \varepsilon_{min}$=3.5, Fig.\ref{stabilized cycles}(a),while the macroscopic cycle is almost elastic for the smallest strain range $\Delta \varepsilon/\Delta \varepsilon_{min}$=1. The detailed experimental cyclic behavior of this material can be found in \citep{Cruzado2017}.

\section{Computational homogenization framework}

The evolution of the stress and strain fields in the microstructure of the polycrystal is obtained by means of the numerical simulation of the stable cyclic response of a Representative Volume Element (RVE) of the microstructure. Simulations are carried out using the finite element method and the elasto-plastic mechanical behavior of each grain within the polycrystal is approximated using a crystal-plasticity model. The FIPs that dictate the number of cycles to initiate a crack in each point of the microstructure are obtained from the simulations. The different parts of the computational homogenization strategy are described below.

\subsection{RVE and finite element model}

\begin{figure}[H]
\centering
{\includegraphics[width=0.5\textwidth]{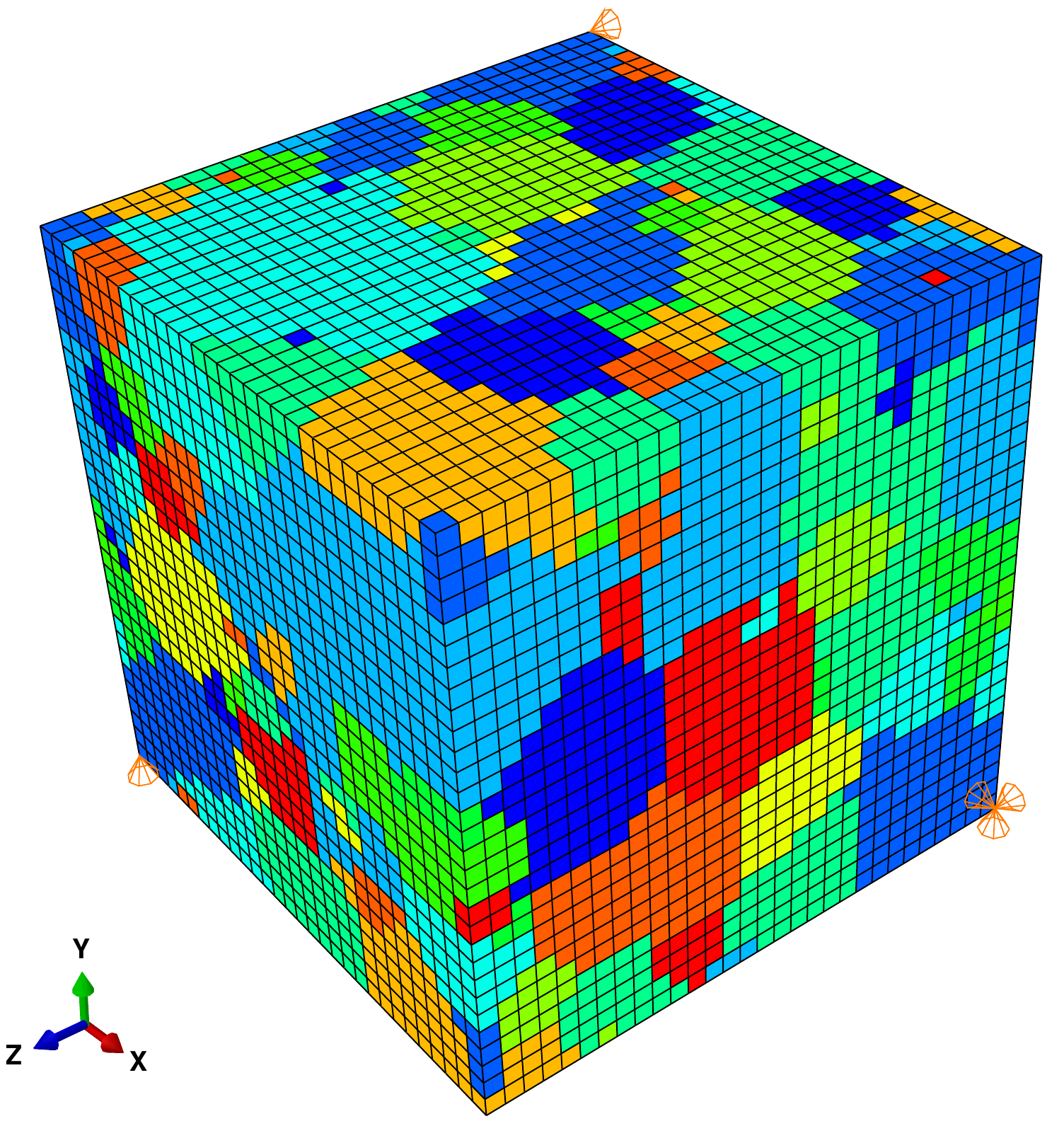}}
\caption{\footnotesize Finite element discretization of an RVE of the microstructure of IN718 alloy with an average ASTM grain size of 8.5.}  
\label{Fig_2:Polycrystal}
\end{figure}

The simulation of the polycrystal behavior was performed using a cubic RVE. The grain size distribution followed a lognormal function whose parameters agree with those corresponding to an ASTM grain size of 8.5. Grains were equiaxied and the grain size distribution was generated using the software Dream3D \cite{Dream3D}. Each grain was discretized with linear brick elements (C3D8 elements in Abaqus) and a typical RVE of the polycrystal is depicted in Figure \ref{Fig_2:Polycrystal}. The minimum size of the RVE to obtain an effective cyclic response of the polycrystal independent of the RVE size, was determined in a previous investigation \cite{Cruzado2017}, leading to 300 grains and 90 elements per grain. This corresponds to a RVE with an approximate size of 109x109x109 $\mu$m for the microstructure studied in this work.

Periodic boundary conditions were applied on opposite faces of the RVE. Strain-controlled uniaxial cyclic deformation was introduced by imposing a cyclic displacement to the master node that controls the deformation of two opposite faces, while the displacements were free in the two perpendicular orientations.  Simulations were carried out using Abaqus/Standard and more details about them can be found in \cite{Cruzado2017}.

\subsection{Crystal plasticity model for cyclic deformation}

The cyclic behaviour of IN718 alloy was simulated with a phenomenological crystal plasticity model recently developed \cite{Cruzado2017}. The main characteristics of the cyclic deformation of IN718 alloy (Bauschinger effect, mean stress relaxation and cyclic softening) are accurately captured by the model \cite{Cruzado2017, Gustafsson2011}. The model description and implementation is described in detail in \cite{Cruzado2017} but the main features are briefly outlined here for completeness. 

IN718 alloy crystals are assumed to behave as an elasto-viscoplastic solids. Plastic deformation occurs in the FCC $\gamma$ phase along 12 slip systems corresponding to the \{111\}   $\textless$110$\textgreater$ family. The plastic slip rate in each slip system, $\dot{\gamma}^\alpha$ depends on the resolved shear stress on the slip system, $\tau^\alpha$, following a power-law, and it is expressed as,

\begin{equation}
\label{eq_(1):gamma_dot}
\dot{\gamma}^\alpha=\dot{\gamma}_0 \left(
  \frac{| \tau^\alpha-\chi^\alpha |}{g^\alpha}\right)^\frac{1}{m}\mathrm{sign} (\tau^\alpha-\chi^\alpha)  
\end{equation}

\noindent where $\dot{\gamma}_0$ is the reference strain rate, ${m}$ the rate sensitivity parameter, ${g^\alpha}$ the critical resolved shear stress (CRSS) on the ${\alpha}$ slip system and $\chi^\alpha$ the back stress. The evolution of $\chi^\alpha$, that determines the kinematic hardening contribution, follows the modified version of the Ohno and Wang model proposed in \cite{Cruzado2017}, 

\begin{equation}
\label{eq_(2):kinematic_OWM}
\dot{\chi}^\alpha=c\dot{\gamma}^\alpha-d\chi^\alpha|\dot{\gamma}^\alpha|\left( \frac{|\chi^\alpha|}{c/d}\right)^{k}   
\end{equation}

\noindent where $c$, $d$ and $k$ are three material parameters. 

The CRSS ${g^\alpha}$ in the model has two contributions, $g^\alpha_m$ and  ${g}_c$,  that arise from the monotonic and cyclic deformation. $\dot g^\alpha_m$  depends on the the accumulated plastic slip, is negative (isotropic softening) and follows the Asaro-Needleman model \cite{Asaro1985} with three parameters that stand for the initial CRSS, $\tau_0$, the saturation CRSS, $\tau_s$, and the initial hardening modulus, $h_0$. The cyclic contribution is also negative (in accordance with the cyclic softening of this alloy) and evolves with the accumulated cyclic plastic strain following a Voce-type law \citep{Tome1984} defined by three parameters, $\tau_s^{cyc}$, $h_1$ and $h_2$, as described in \cite{Cruzado2017}. 

The parameters of the model to simulate  the cyclic deformation of IN718 alloy  at 400$^\circ$C  are presented in Table \ref{Table:CP_parameters}. The elastic constants were obtained from the experimental values reported by \cite{Martin2014} at ambient temperature assuming a linear reduction of the three elastic constants with temperature similar to the one found for the elastic modulus of IN718 alloy polycrystals. The reference strain rate, $\dot{\gamma}_0$, strain rate sensitivity exponent, $m$, and latent hardening coefficients $q_{\alpha\beta}$ were obtained from mechanical tests on single crystal micropillars machined from the polycrystals \citep{Cruzado2015}. The remaining parameters of the crystal plasticity model were obtained in \cite{Cruzado2017} using the inverse optimization approach proposed in \citep{HerreraSolaz2015} from the experimental results of the cyclic response of IN718 alloy at 400$^{\circ}$C under $R_{\varepsilon}$ = 0. Note that the viscoplastic parameters are normalized by the initial critical resolved shear stress $\tau_0$.

\begin{table}[H]
\centering
\begin{tabular}[5pt]{lcccc}
\hline\noalign{\vskip2pt}
\multirow{2}{*}{Elastic}
  &C$_{11}$(GPa)  &   C$_{12}$(GPa)   &   C$_{44}$(GPa)    \\[0.2cm]
& 240 & 165 & 101  \\
\hline\noalign{\vskip2pt}
\multirow{2}{*}{Viscoplastic}
 &  $m$ & $\dot{\gamma}_0$ &  &      \\[0.2cm]
 & 0.017 &    2.42 $10^{-3}$   &      &          \\[0.5cm]
 \hline\noalign{\vskip2pt}
 \multirow{2}{*}{Isotropic softening} & $\tau_0$ (MPa)  & $\tau_s$ & $h_0$  & $q_{\alpha\beta}$ \\[0.2cm]
& $\tau_0$ & 0.71$\tau_0$ & 57.13$\tau_0$ & 1 \\[0.5cm]
\hline\noalign{\vskip2pt}
\multirow{2}{*}{Kinematic hardening}  & $c$ & $d$ & $mk$  \\[0.2cm]
& 58.9$\tau_0$ & 198.3 & 17.7  \\[0.5cm]
\hline\noalign{\vskip2pt}
\multirow{2}{*}{Cyclic softening} & $\tau_s^{cyc}$ & $h_{1}$ & $h_2$ &   \\[0.2cm]
  & 0.076$\tau_0$ & 0.07$\tau_0$ & 2.33 $10^{-6}$  $\tau_0$ &\\[0.5cm]
\noalign{\vskip2pt}\hline
\end{tabular}
\caption{Crystal plasticity parameters for a wrought IN718 alloy at $400^{\circ}$C.}
\label{Table:CP_parameters}
\end{table}

\subsection{Local fatigue indicator parameters}

The local Fatigue Indicators Parameters (FIPs) provide information about the evolution of the stress and strain fields in each fatigue cycle under steady-state conditions. They vary across the RVE depending on the local features of the microstructure and can be related to the number of cycles necessary to nucleate a crack. Different FIPs have been reported in the literature to describe  the main driving force that controls crack initiation. They include the accumulated plastic slip per cycle \citep{Manonukul2004, Sweeney2012, Sweeney2014, Sweeney2015}, the Fatemi-Socie parameter \citep{Shenoy2007, Musinski2012,Przybyla2013,Castelluccio2013, Castelluccio2014} and the energy stored or dissipated per cycle during cyclic deformation \citep{Sweeney2012, Sweeney2014, Sweeney2015, Wan2014, Wan2016}. The latter one will be used in this investigation, which assumes that crack incubation is driven by the strain energy dissipated in each particular slip plane because persistent slip band formation is associated to a specific crystallographic planes. 

The local crystallographic strain energy dissipated per cycle can be expressed as 

\begin{equation}
W_{cyc}^\alpha(\mathbf{x})= \int_{cyc} \tau_\alpha(\mathbf{x})\dot{\gamma}_\alpha(\mathbf{x}) \mathrm{d}t
\label{eq_(4):W}
\end{equation}\\[-0.5cm]

\noindent where $\tau_\alpha$ and $\dot{\gamma}_\alpha$ are the resolved shear stress and the shear strain rate on the slip system $\alpha$, respectively. The FIP is calculated in  the centroid of each element in the RVE (whose coordinates are given by $\mathbf{x}$), but this local value depends on the details of the discretization and it is not representative of a fatigue damaged region  \citep{Sweeney2013}. From a physical viewpoint, the FIP should be averaged over a region representative of the crack incubation zone, while  the volume averaging is indeed necessary to avoid spurious stress concentrations and minimize mesh size effects from the computational perspective, as noted in \citep{Castelluccio2015}.

An averaging approach, similar to the one proposed by Castelluccio {\it et al.} \citep{Castelluccio2015},  is used in this investigation. Four different bands of thickness $h$ (equal to the diagonal of the voxel) are defined in the centroid of each element, parallel to each one the slip planes of the FCC crystal. This band averaging is able to capture the anisotropic  deformation in each point of the RVE.  Therefore, it can account for the localization of plastic deformation in bands parallel to the slip planes, that give rise to the formation of persistent slip bands. Under these assumptions, the local FIP representative of an RVE subjected to cyclic deformation, $W_{cyc}^b$, is obtained as the maximum of the band-averaged  local FIP throughout the RVE, according to

\begin{eqnarray}
W_{cyc}^b=\max_{i=1,nb} \left \{\max_{\beta_i} \frac{1}{V_i}\int_{V_i} W_{cyc}^{\beta_i}(\mathbf{x})\mathrm{d}V_i \right \}
\label{eq_(5):FIPfinal}
\end{eqnarray}

\noindent  where $\beta_i$ (= 1, 2, 3) corresponds to the three different slips systems contained in the slip plane parallel to the band $i$, $V_i$ is the volume of that band and $nb$ is the total number of bands in the microstructure, which is 4 times the number of elements in the RVE.

\section{Fatigue life prediction} 

The computational homogenization framework proposed in the previous section is used here to predict the low-cycle fatigue behavior of IN718 alloy. To this end, the band-averaged, local FIP in the RVE,  $W_{cyc}^b$, is determined as indicated in section 4.1. Based on the hypothesis that the crack initiation is transgranular in both C-M regimes and is consequence of the different plasticity behavior generated at the different strain ranges as reported in \citep{Praveen2008,Praveen2004} the FIP is used in section 4.2 to develop a microstructure-based fatigue lifemodel able to capture the the bilinear C-M behavior. 
Finally the predictive capabilities of the model for different strain ranges and strain ratios are discussed  in section 4.3.

\subsection{Cyclic evolution of the FIP fields}

Fatigue crack nucleation is related to the stabilized value of the cyclic FIP ($W_{cyc}^b$) given by eq. (\ref{eq_(5):FIPfinal}), which is determined from the numerical simulation of the cyclic behavior of an RVE. Most of the micromechanics-based fatigue life models in the literature assume that the stabilization of the cyclic stress-strain behavior (and, thus, of $W_{cyc}^b$) is reached after a few cycles \citep{Sweeney2015, Wan2016, Shenoy2007, Castelluccio2015}. This is not, however, the case for IN718 alloy because cyclic softening or mean stress relaxation could extend the initial transient period until  stabilization up to a significant fraction of the fatigue life. Simulation of the cyclic behavior for hundreds or thousands of cycles is very expensive from the computational viewpoint  and cyclic jump approaches have been developed to reduce the computational cost \citep{Joseph2010,Ghosh2011, Cruzado2012}. In this investigation, the cyclic jump strategy proposed by Cruzado {\it et al.} \citep{Cruzado2017},  based  on the linear extrapolation of the evolution of the internal variables at the constitutive equation level, followed by one or several stabilization cycles, was used. The detailed description of this approach as well as the parameters used for the cycle jumps are described in detail in \cite{Cruzado2017}.

\begin{figure}[!]
\includegraphics[scale=0.90]{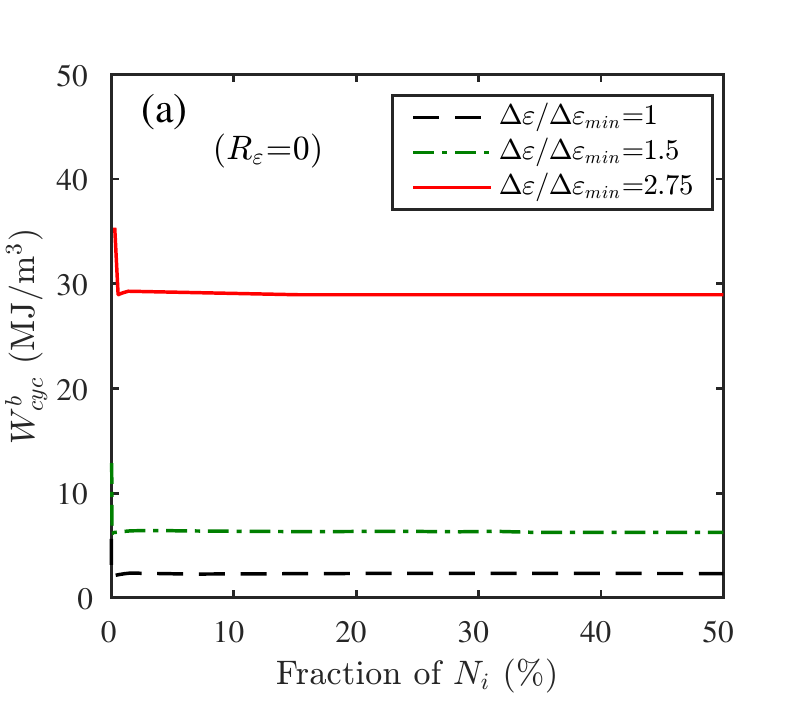}
\includegraphics[scale=0.90]{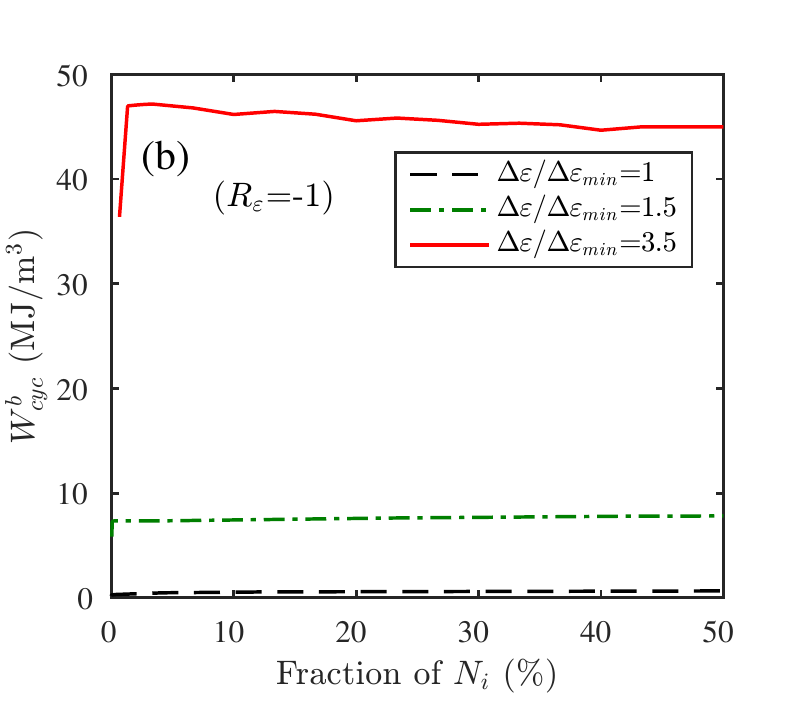}
\includegraphics[scale=0.90]{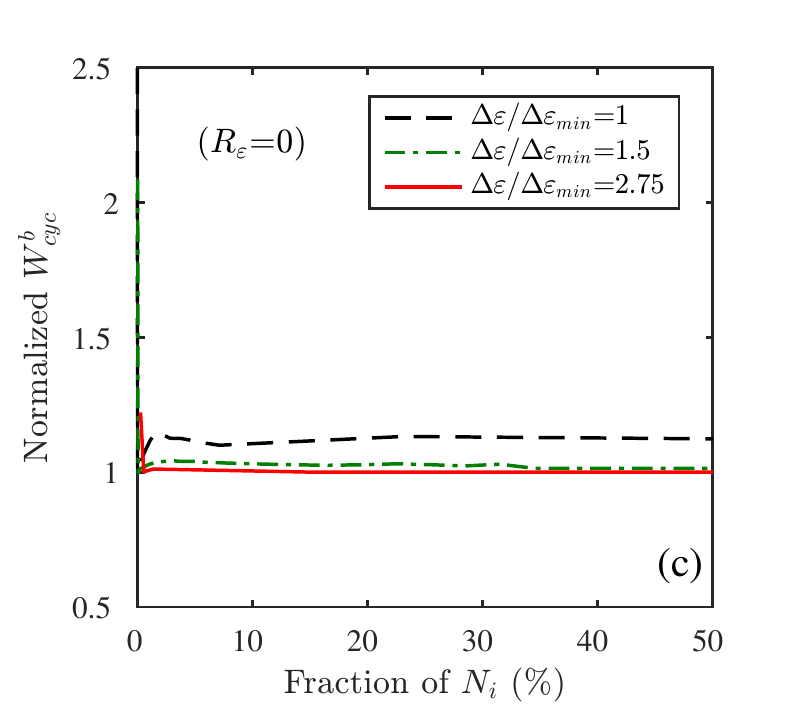}
\includegraphics[scale=0.90]{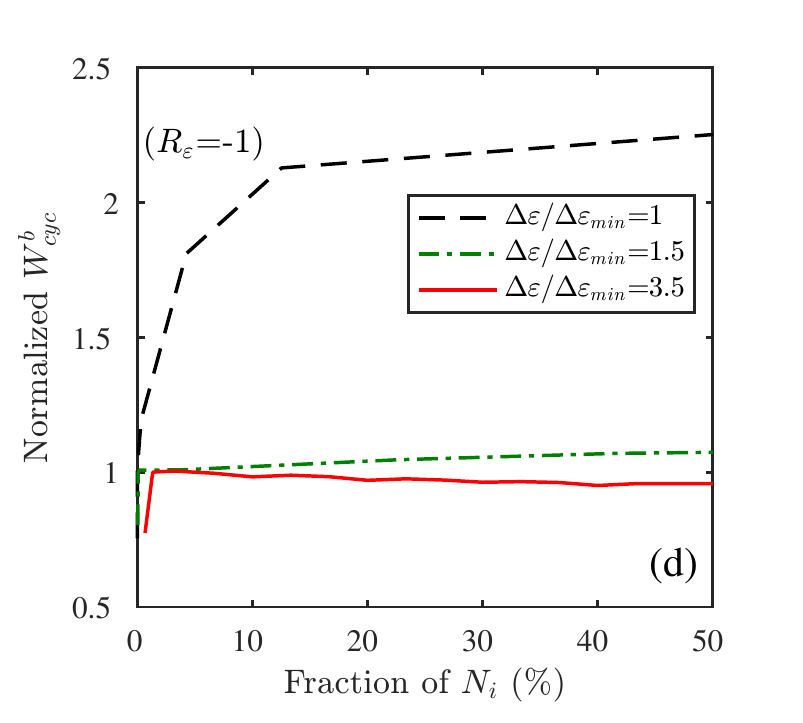}
\caption{Evolution of the band-averaged, cyclic strain energy dissipation FIP, $W_{cyc}^{b}$, as a function of the fraction of cycles to crack nucleation, $N_i$, for different strain ranges. (a)  Strain ratio $R_\varepsilon$=0. (b) Strain ratio $R_\varepsilon$=-1. (c) {\it Idem} as (a) but $W_{cyc}^{b}$  for each strain range is normalized by the corresponding value of $W_{cyc}^{b}$ in the second cycle. (d) {\it Idem} as (b) but $W_{cyc}^{b}$  for each strain range is normalized by the corresponding value of $W_{cyc}^{b}$ in the second cycle.}
\label{Fig_3:FIP_evolution}
\end{figure}

Using these strategies, the evolution of $W_{cyc}^b$ is plotted in Fig.\ref{Fig_3:FIP_evolution}(a) and (b) as a function of the fraction of cycles for crack initiation, $N_i$,  for two different strain ratios, $R_\varepsilon$=0 and $R_\varepsilon$=-1, respectively and various cyclic strain ranges for each strain ratio. $W_{cyc}^b$ decreased significantly in the case of $R_\varepsilon$=0 from the first to the second cycle as a  consequence of the reduction in the plastic strain accumulated between the first and the second cycle. Afterwards, $W_{cyc}^b$ remained almost constant and differences in the first five cycles can only be appreciated in  Fig \ref{Fig_3:FIP_evolution}(c), in which the evolution of $W_{cyc}^b$ with the number of cycles is normalized by the value of $W_{cyc}^b$ in the second cycle. Thus, stable values of $W_{cyc}^b$ can be obtained by simulating 5 loading cycles in the case  of $R_\varepsilon$=0.

However, $W_{cyc}^b$ increased from the first cycle under fully-reversed cyclic deformation, $R_e$=-1, Fig \ref{Fig_3:FIP_evolution}(b), as a result of the increase in the plastic strain accumulated during the initial tensile loading step and in the subsequent compression-tension steps. A stable value of  $W_{cyc}^b$ was only reached after 50\% of the fatigue life, particularly for small cyclic strain ranges ($\Delta\varepsilon/\Delta\varepsilon_{min}$ =1.0). This behavior is clearly shown if the the evolution of $W_{cyc}^b$  with the number of cycles for each applied cyclic strain range is normalized by the corresponding value of $W_{cyc}^b$ after the second cycle, Fig \ref{Fig_3:FIP_evolution}(d). While the FIP remained constant after a few cycles in the case of $R_\varepsilon$=0, $W_{cyc}^b$ increased rapidly during the initial 12\% of the fatigue life for $R_\varepsilon$= -1 and $\Delta\varepsilon/\Delta\varepsilon_{min}$ =1.0. It should be noticed that the macroscopic cyclic response is almost elastic and plastic deformation occurs at the microlevel only in a few favourably oriented grains under these conditions ($R_\varepsilon$= -1 and $\Delta\varepsilon/\Delta\varepsilon_{min}$ =1.0). The cyclic plastic deformation in these grains is small, $\Delta \gamma<$ 0.014\%  but grows slowly and continuously upon cyclic loading due to two mechanisms. The first one is the development of cyclic softening, which is included in the constitutive model \cite{Cruzado2017}. The second one is the accommodation of the elastic strains  in the neighbor grains until the steady state is attained.  Thus, plastic strains and the corresponding FIP  are progressively localized  for small $\Delta\varepsilon$ in a few grains, which control the number of cycles for fatigue crack nucleation.

It is obvious from the previous analysis that the simulation of the cyclic deformation of the RVEs has to be extended up to the number of cycles at which the cyclic FIP becomes stable. This occurs in a few cycles in the case of $R_\varepsilon$= 0 or $R_\varepsilon$= -1 and large cyclic strain ranges. However, a large number of cycles (up to 50\% of the fatigue life) has to be simulated in the case of $R_\varepsilon$= -1 and small cyclic strain ranges. Otherwise, large errors will be introduced in the values of the stabilized cyclic FIP. Thus, all the simulations presented below were carried out for the first 5 cycles except for the case of $\Delta\varepsilon/\Delta\varepsilon_{min}$ =1.0 and $R_\varepsilon$= -1, in which they are extended up to 12\% of the fatigue life. The differences in $W_{cyc}^b$ between 12\% and 50\% of the fatigue life were small and did not justify the huge computational effort to extend the simulations up to 50\% of the fatigue life

\subsection{Microstructure-based fatigue life model}

\begin{figure}[H]
\centering
\includegraphics[scale=0.8]{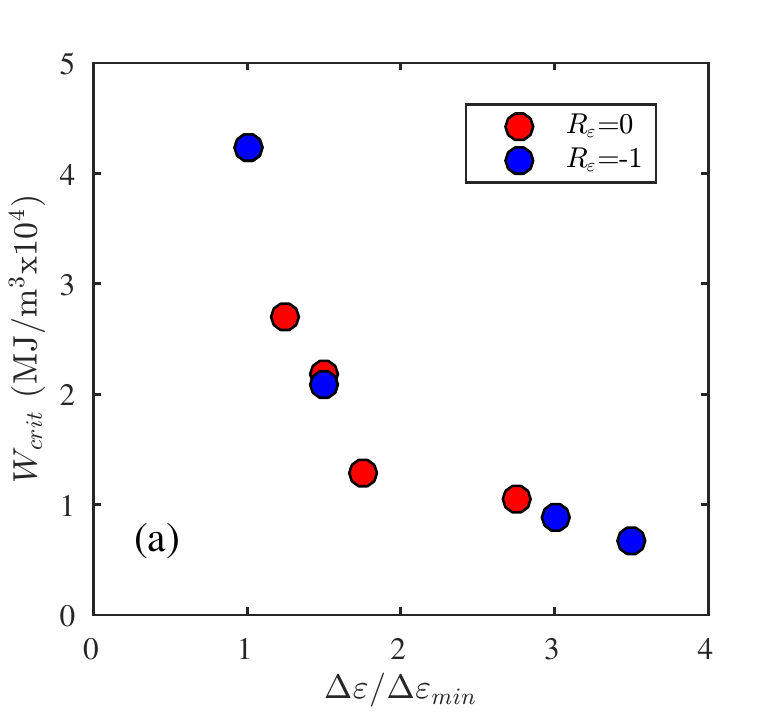}
\includegraphics[scale=0.8]{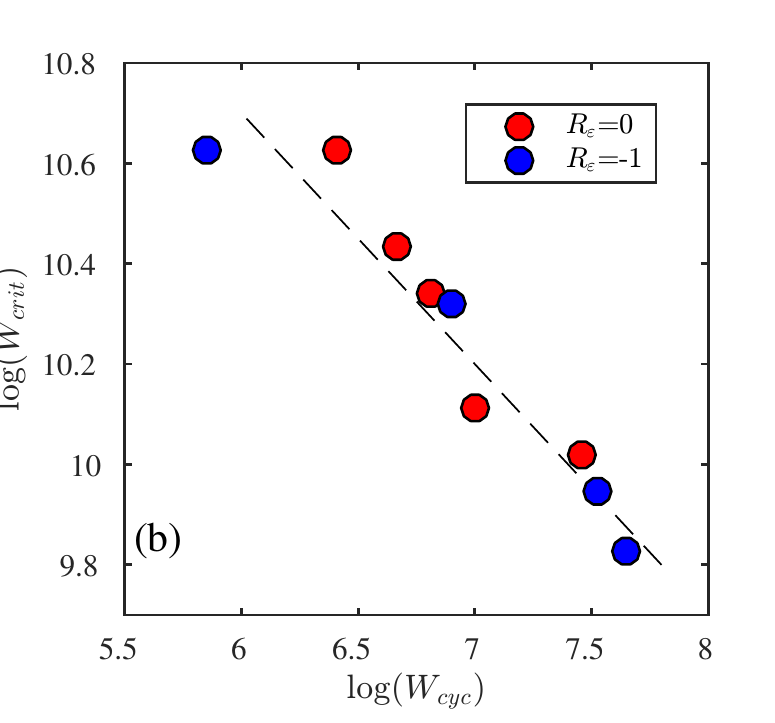}
\caption{(a) Evolution of the critical FIP, $W_{crit}$, as given by the linear model in Eq.(\ref{eq_(9):FIP_crit}), as a function of the cyclic strain range, $\Delta\varepsilon$ and strain ratio, $R_\varepsilon$ in IN718 alloy at 400 $^{\circ}$C. (b) Value of the the critical FIP, $W_{crit}$, as function of the $W_{cyc}^b$ obtained when simulating the cyclic response under the different cyclic strain ranges.} 
\label{Fig_4:Plastic_vs_wcrit}
\end{figure}

Fatigue life models in the literature are based on a critical value of the accumulated cyclic FIP  \citep{Manonukul2004, Sweeney2015, Wan2014} at which crack initiation is produced. This critical value, $W_{crit}$,  can be  obtained from computational homogenization simulations and one experiment using a simple linear model. The number of cycles for fatigue crack initiation (or the fatigue life), $N_i$, is measured in a test for a particular  cyclic strain range and strain ratio  and the corresponding stabilized cyclic FIP, $W_{cyc}^b$, is provided by the simulations of the cyclic behavior of the RVE under the same loading conditions. The fatigue life, $N_i$, is then calculated as 

\begin{equation}
N_i = \frac{W_{crit}} {W_{cyc}^b}
\label{eq_(9):FIP_crit}
\end{equation}

\noindent where $W_{crit}$ is a material constant and can be used to predict the fatigue life under different loading conditions. This assumption is valid, as demonstrated by Manonokul and Dunne \citep{Manonukul2004}, for materials that exhibit a linear C-M relationship. This is not, however, the case in materials that exhibit a bilinear C-M response. This is illustrated in Fig. \ref{Fig_4:Plastic_vs_wcrit} in which $W_{crit}$ (obtained from the experimental values of $N_i$ and the numerical predictions for $W^b_{cyc}$ in eq. (\ref{eq_(9):FIP_crit})) is plotted for different cyclic strain ranges, $\Delta\varepsilon$, and strain ratios, $R_{\varepsilon}$ in IN718 alloy. It is obvious that 
$W_{crit}$ is not constant  and increases rapidly as  $\Delta\varepsilon$ decreases.

The behavior depicted in Fig. \ref{Fig_4:Plastic_vs_wcrit} implies that $W_{crit}$ is not a material parameter but depends also on the loading conditions and can be attributed to the change in deformation mechanism that controls the nucleation of fatigue cracks: from localized deformation in a few grain at small cyclic strain ranges to homogeneous plastic deformation at large cyclic strain ranges. Furthermore, it was observed that the value of $W_{crit}$ presents a doble logarithmic relation with the FIP value, as it can be observed in Fig. \ref{Fig_4:Plastic_vs_wcrit}(b). If this relation between $W_{crit}$ and $W_{cyc}$ is introduced in eq. \ref{eq_(9):FIP_crit} the expression relating the fatigue life $N_i$ with $W_{cyc}$  turns into a power-law relation

\begin{equation}
N_i= \frac{W_{crit}^{NL}} {(W_{cyc}^b)^m}
\label{eq_(11):energy}
\end{equation}

\noindent in which the fatigue life depends on the stabilized cyclic FIP, $W_{cyc}^b$, through two material parameters, $W_{crit}^{NL}$ and $m$. Obviously, $W_{crit}^{NL} = W_{crit}$ if $m$ = 1 and the non-linear model is reduced to the linear one in this case. Of course, two experimental fatigue tests (instead of one) are necessary to obtain the parameters $m$ and $W_{crit}^{NL}$ of the non-linear fatigue life model and this is consistent with the underlying hypothesis: the bilinear C-M relationship is consequence of the differences in the distribution of deformation between small and large cyclic strain ranges corroborating the hypothesis made previously.   

\subsection{Results and discussion}

The parameters of the microstructure-based non-linear fatigue life model (eq. (\ref{eq_(11):energy}) were obtained from the experimental results corresponding to applied cyclic strain ranges of $\Delta \varepsilon /\Delta \varepsilon_{min}$ = 1 and 3.5 and a strain ratio $R_\varepsilon$=-1. It should be noted, however, that the number of grains in the RVE is always much smaller than the number of grains in the actual microstructure and the values of the FIP obtained in the simulation of different RVE corresponding to the same microstructure can be quite different. Thus, a single model is not fully representative of the fatigue life for a given microstructure and does not provide reliable information about the scatter resulting from the particular orientation arrangements. To overcome these limitations, the stabilized cyclic FIP, $W^b_{cyc}$,  was obtained for each loading condition by aggregating the results obtained with 20 different realizations of the RVE named SVE (Statistical Volume Elements), \citep{Shenoy2006, Castelluccio2013}. The value of the FIPs ($W^b_{cyc}$) obtained with the 20 RVEs for each of the two loading conditions was used to adjust the fatigue nucleation expression to a normal distribution and a good agreement was found. In the case of $\Delta \varepsilon /\Delta \varepsilon_{min}$ = 3.5, the standard deviation found was only 10\% of the FIP average value, while for the smallest deformation a much larger value of 58\% of the mean was obtained. The mean value of the FIPs distributions were used to obtain the parameters of the non-linear fatigue life and the resulting values were $W_{crit}^{NL}$ = 6.3770 x 10$^4$ MJ/m$^3$ and $m$ = 1.5641.

\begin{figure}[H]
\includegraphics[scale=0.90]{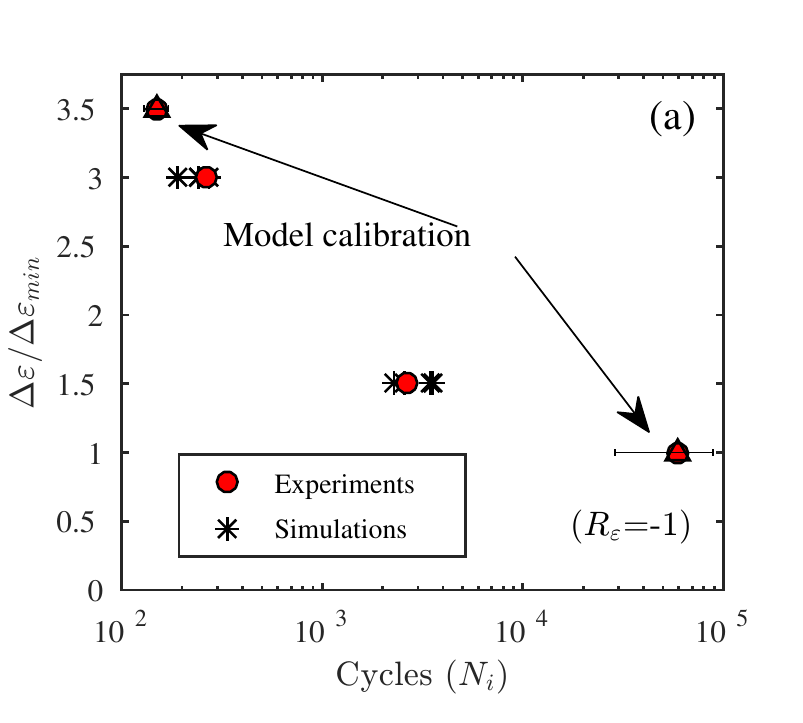}
\includegraphics[scale=0.90]{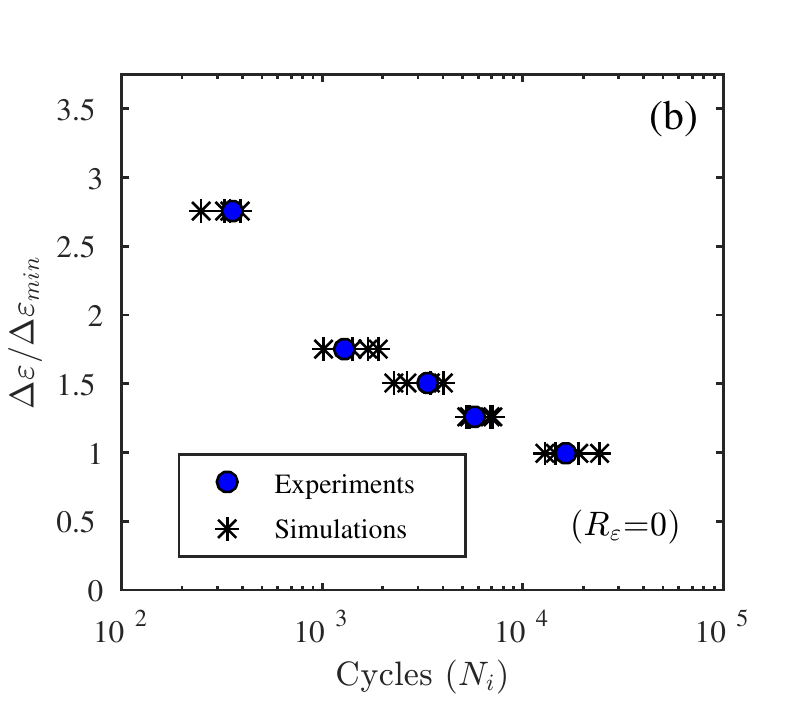}
\caption{Experimental  results and model predictions for the fatigue life of IN718 alloy at 400$^{\circ}$C.
 (a) $R_{\varepsilon}$ = -1 and  (b) $R_{\varepsilon}$ = 0}
\label{Fig_5:Fatigue Life predictions}
\end{figure}

The ability of the proposed  non-linear model to predict the fatigue life of IN718 alloy at 400$^\circ$ is depicted in Fig. \ref{Fig_5:Fatigue Life predictions}(a) and (b), in which the experimental and the model results for different cyclic strain ranges are plotted  for tests carried out with $R_\varepsilon$ = -1 and 0, respectively. The model predictions were carried out with four different RVEs, that are within the standard deviation obtained from the 20 RVEs to have a statistical representation similar to the one obtained in the large sample for each loading case and so, four different numerical results are plotted in Fig. \ref{Fig_5:Fatigue Life predictions} for each experimental point. In the case of $\Delta \varepsilon /\Delta \varepsilon_{min}$ = 1 and 3.5 and $R_\varepsilon$=-1, the numerical results are represented by a horizontal line that includes the data obtained from the simulation of 20 realizations of the RVEs (the ones used to determine $W_{crit}^{NL}$ and $m$).  The agreement between experiments and simulations was excellent in all cases and the simulation strategy was also able to capture the life reduction induced by $R_{\varepsilon}$=0, as compared with the tests carried out with $R_{\varepsilon}$ = -1.

\begin{figure}[H]
\includegraphics[scale=0.90]{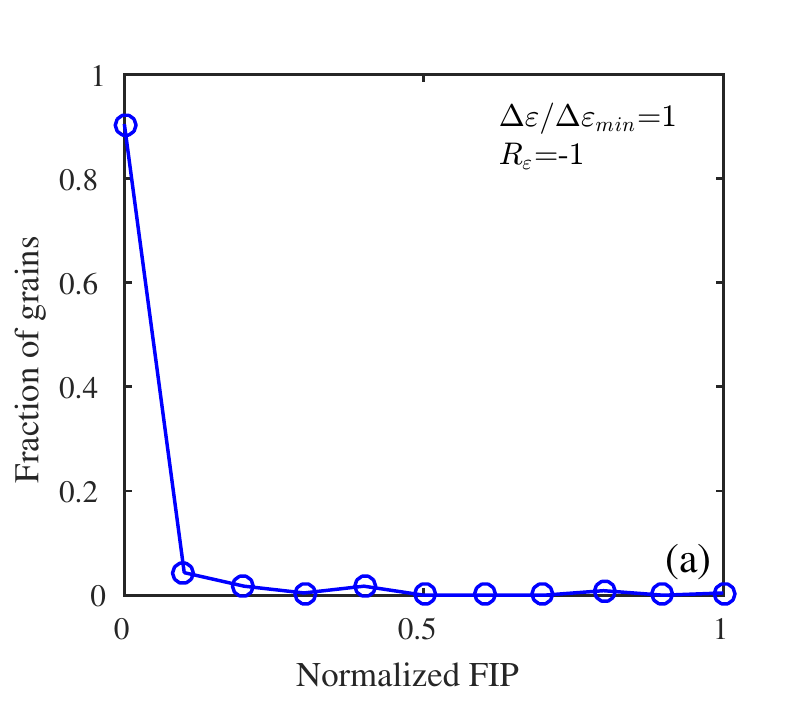}
\includegraphics[scale=0.90]{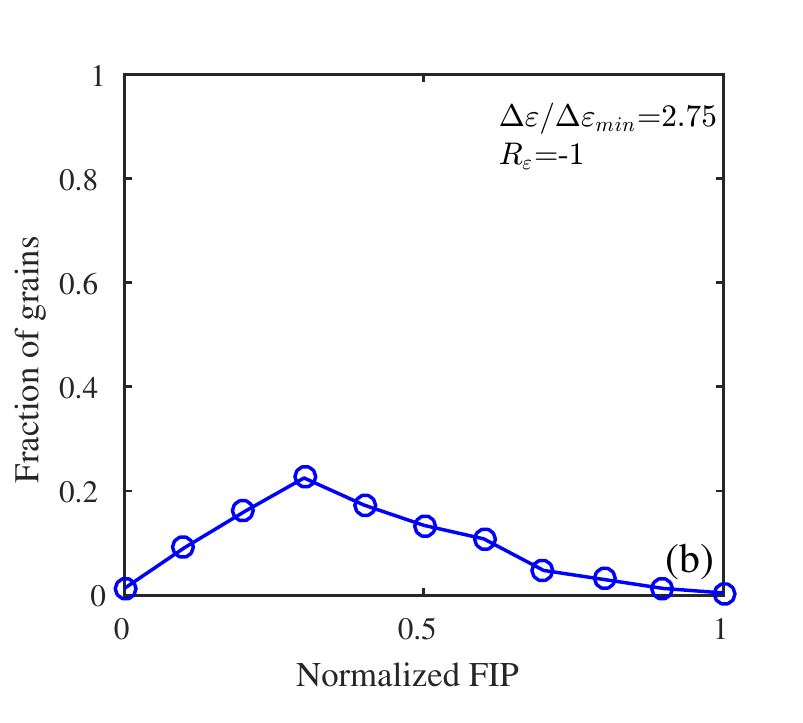}
\caption{Histogram of the stabilized cyclic FIP (normalized with respect to the maximum value of the corresponding FIP, $W_{cyc}^{b}$) at a strain ratio $R_\varepsilon$ = -1. (a) $\Delta \varepsilon /\Delta \varepsilon_{min}$ =1 for one SVE. (b)  $\Delta \varepsilon /\Delta \varepsilon_{min}$ = 2.75.} 
\label{Fig_6:Distribution FIPs}
\end{figure}

The model is able to predict accurately the fatigue life under small and large cyclic strain ranges because it can capture the differences in the microfield distributions at the different  strain ranges. This is demonstrated in Figs. \ref{Fig_6:Distribution FIPs}(a) and (b), which shows the histograms of the stabilized, cyclic FIP in each grain (normalized by the maximum value in the whole SVE) for the  two extreme cyclic strain ranges of $\Delta\varepsilon/\Delta\varepsilon_{min}$=1 and 2.75, respectively,  and $R_\varepsilon$=0. In the case of $\Delta\varepsilon/\Delta\varepsilon_{min}$ = 1, the stabilized, cyclic FIP in 90\% of the grains is zero and the fatigue life is controlled by a very small population of grains in which plastic deformation is localized. As a result, the scatter in fatigue life is large and this is shown in the scatter bar for the fatigue life predictions for this loading conditions  in Fig. \ref{Fig_5:Fatigue Life predictions}(a). On the contrary, the distribution of the stabilized, cyclic FIP in the case of $\Delta\varepsilon/\Delta\varepsilon_{min}$ = 2.75 shows that plastic strain and energy dissipation occur in most of the grains and the differences between the maximum and minimum stabilized, cyclic FIP is below 2 for 50\% of the grains. Thus, plastic deformation is rather homogeneous throughout the microstructure and the scatter induced by the microstructure of the fatigue life is reduced,  as shown by  the corresponding scatter bar in Fig. \ref{Fig_5:Fatigue Life predictions}(a).
For strain amplitudes lower than the ones studied in this work, where the fatigue life enters in the high cycle fatigue  regime, other microstructural details such as twin boundaries \citep{Yeratapally2016}, precipitates \citep{Sweeney2014} and free surfaces \citep{Wan2016,Sweeney2014, Przybyla2012} may play an important role. 
These microstructural defects act as stress concentrators, leading to localized plastic microfields in the vicinity of the defects that will be probable sites for the nucleation of persistent slip bands and small cracks. The consideration of these defects will not change very much the present results, but they should definitively be accounted for when predicting the life at lower strain amplitudes. 

\begin{figure}[H]
\includegraphics[scale=0.83]{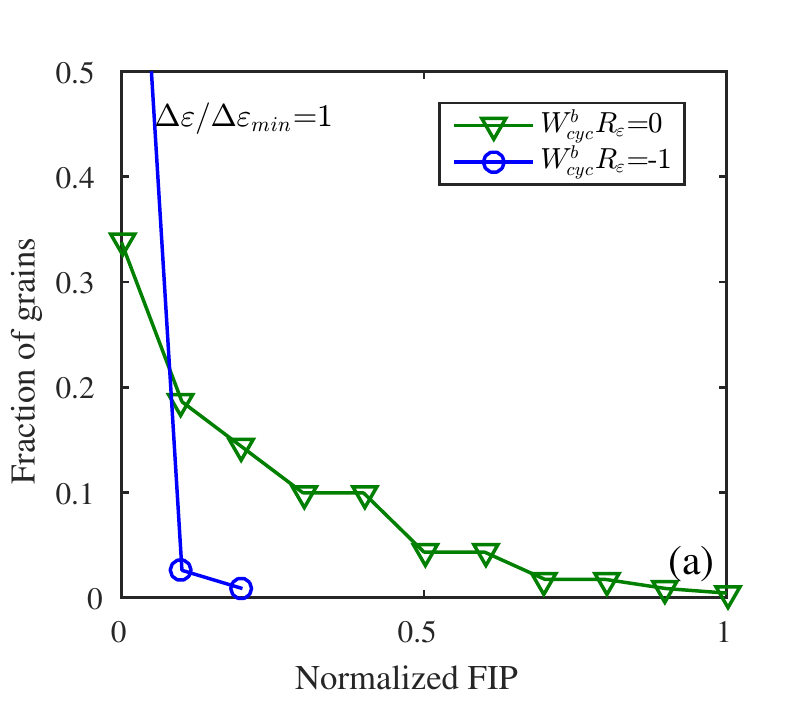}
\includegraphics[scale=0.83]{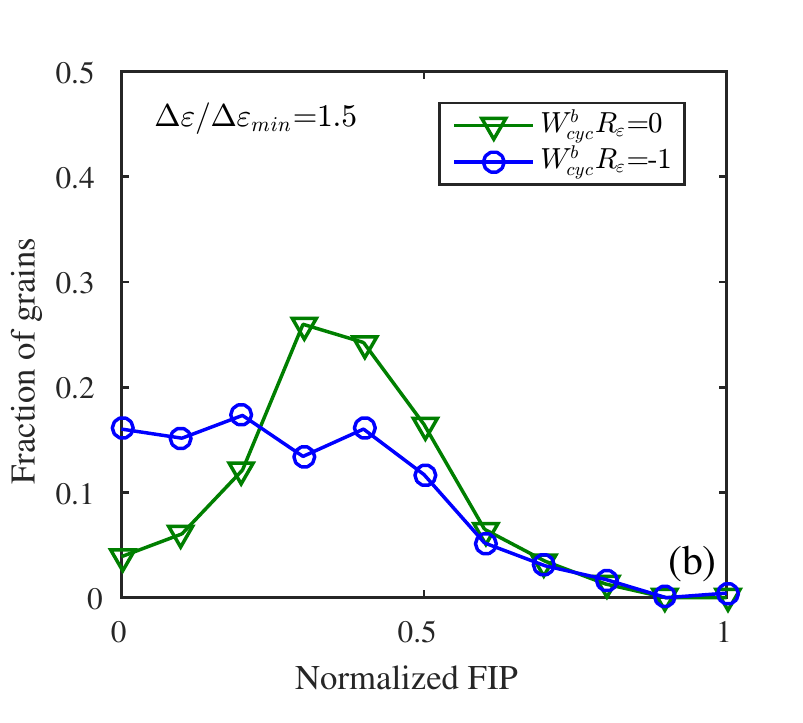}
\centering
\includegraphics[scale=0.83]{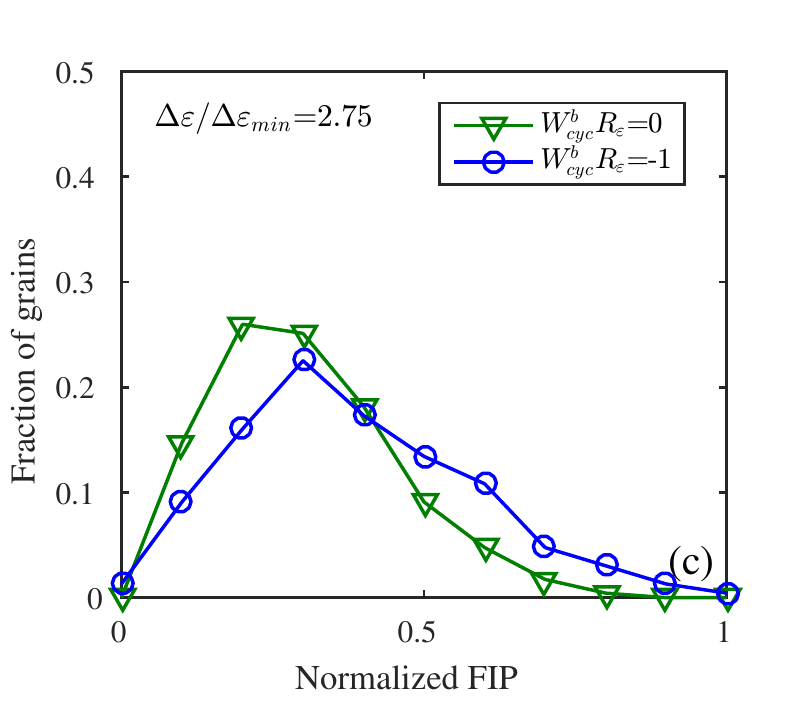}
\caption{Histogram of the stabilized cyclic FIP in one RVE for two different values of the strain ratio, $R_\varepsilon$=-1 and $R_\varepsilon$=0. a) Cyclic strain range $\Delta \varepsilon /\Delta \varepsilon_{min}$ =1.  (b) Cyclic strain range $\Delta \varepsilon /\Delta \varepsilon_{min}$ =1.5.  (c) Cyclic strain range $\Delta \varepsilon /\Delta \varepsilon_{min}$ =2.75. The FIP values in each figure are normalized by the maximum value of the FIP in the simulations carried out with either $R_\varepsilon$ = 0 or $R_\varepsilon$ = -1.} 
\label{Fig_7:FIP_distribution_strain_ratio}
\end{figure}

The experimental results of IN718 alloy at 400$^\circ$C showed that the strain ratio has very little influence on the  fatigue life for large strain ranges while asymmetric cycles clearly accelerate fatigue crack nucleation and reduce the fatigue life at low cyclic strain ranges (Fig. \ref{Fig_5:Fatigue Life predictions}). The model was able to capture this behavior and to provide an explanation at the micromechanical level. The histograms of the distribution of the stabilized, cyclic FIP within the RVE are plotted in Figs. \ref{Fig_7:FIP_distribution_strain_ratio}(a), (b) and (c) for cyclic strain ranges of $\Delta\varepsilon/\Delta\varepsilon_{min}$ = 1.0, 1,2 and 2.75, respectively, and two different values of the strain ratio, $R_\varepsilon$ = 0 and $R_\varepsilon$ = -1. The FIP distributions in each figure are normalized by the maximum value of the FIP in the simulations carried out with either $R_\varepsilon$ = 0 or $R_\varepsilon$ = -1 to highlight the influence of this parameter on the fatigue life predictions.
 The FIP distribution throughout the microstructure is very different in the cases of $R_\varepsilon$=0 and $R_\varepsilon$ = -1 at low cyclic strain range, $\Delta \varepsilon/\Delta\varepsilon_{min}$=1, Fig. \ref{Fig_7:FIP_distribution_strain_ratio} (a).  The maximum value of the FIP ($W_{cyc}^b$) for $R_\varepsilon$=-1 is much smaller than for $R_\varepsilon$=0, in agreement with the experimental results that show much longer fatigue life with $R_\varepsilon$ = -1 for the same cyclic strain range. In addition, most of the grains did not undergo any plastic deformation in the case of $R_\varepsilon$=-1 and were not able to nucleate fatigue cracks. As the cyclic strain range increased, Fig. \ref{Fig_7:FIP_distribution_strain_ratio}(b), the differences in the distribution of the FIPs throughout the microstructure diminished and the maximum value of the FIP, $W_{cyc}^b$, was the same for $R_\varepsilon$=-1 and $R_\varepsilon$=0 at $\Delta \varepsilon/\Delta\varepsilon_{min}$=1.5.  Thus, fatigue life predictions were equivalent for this cyclic strain range although the localization of the strain was more marked for $R_\varepsilon$= -1. Finally, plastic strain was homogeneously distributed across the microstructure for both values of $R_\varepsilon$ at $\Delta \varepsilon/\Delta\varepsilon_{min}$= 2.75, leading to similar fatigue lives that were independent of the local microstructural features.

\section{Conclusions} 

A microstructure-based fatigue life model is proposed for polycrystalline materials that exhibit bilinear Coffin-Manson relationship. The model assumes that the differences in the fatigue behavior in specimens subjected to small and large cyclic strain ranges can be attributed to the transition from highly localized plasticity at low strains to homogeneous deformation at high strain ranges. It was applied and validated for an IN718 alloy subjected to fatigue at 400$^\circ$.

The microstructure-based fatigue life model was based on the analysis of the cyclic deformation of the polycrystal by means of computational homogenization of a representative volume element of the microstructure. The constitutive equation for the single crystals in the polycrystal included the relevant features observed during cyclic deformation of IN718 alloy and the simulations were carried out (using a cyclic jump strategy when necessary) until the plastic deformation was stabilized. The fatigue life was predicted from the maximum value of  a fatigue indicator parameter based of the energy dissipated by plastic deformation in each cycle in the microstructure, $W_{cyc}^b$, which was related to the number of cycles for failure by means of a power law which depended on two material parameters. These material parameters were obtained from the experimental tests carried out at two different cyclic strain ranges ranges (small and large) and $R_\varepsilon$ = -1.

The microstructure-based fatigue life model was able to predict accurately the bilinear Coffin-Manson  behavior  and the fatigue life of IN718 alloy at 400$^\circ$C in a wide range of cyclic strain ranges and different values of the strain ratio, $R_\varepsilon$ = -1 and 0. In agreement with the initial hypothesis, the model showed that the different behavior at small and large cyclic strain ranges was triggered by the localization of the deformation in the former in very few grains in which the plastic strains were slowly increasing during a significant fraction of the fatigue life. In addition, the reduction in the fatigue life in the tests carried out at low strain ranges  with $R_\varepsilon$ = -1, as compared with those carried out with $R_\varepsilon$ = 0, was also successfully explained by the micromechanics model as well as the large experimental scatter found at small cyclic strain ranges. 

\section*{Acknowledgments}

This investigation was supported by the European Union through the Clean Sky Joint Undertaking, 7th Framework Programme, project MICROMECH  (CS-GA-2013-620078) and by the Spanish Ministry of Economy and Competitiveness through the projects DPI2015-67667-C3-2-R and DPI2015-67667-C3-1-R. In addition, the authors thank Industria de TurboPropulsores. S. A. and, in particular,  Dr. A. Linaza and Dr. K. Ostolaza for providing the experimental data and their support during the project. The experimental data included in this paper are proprietary of the industrial partners in the MICROMECH project and, as a result, the stresses and strains have been normalized by a constant factor. Finally, the simulation tools developed in the last stages of this investigation were supported by the European Research Council under the European Union Horizon 2020 research and innovation programme (Advanced Grant VIRMETAL, grant agreement No. 669141).


\end{document}